\documentclass{article}
\usepackage{authblk}
\usepackage[utf8]{inputenc}
\usepackage{diagbox}
\usepackage{amsmath}
\usepackage{amsfonts}
\usepackage{graphicx}
\usepackage{mathtools}
\usepackage{float} 
\usepackage{xcolor}
\usepackage{hyperref}
\usepackage{verbatim}
\usepackage{float}
\usepackage{algorithm} 
\usepackage{algpseudocode}
\providecommand{\keywords}[1]
{
  \small	
  \textbf{\textit{Keywords---}} #1
}
\title{ViscoelasticNet: A physics informed neural network framework for stress discovery and model selection}
\date{}

\author[1]{Sukirt Thakur}

\affil[1]{\small{School of Mechanical Engineering,
            Purdue University, 
            West Lafayette,
            47907, 
            Indiana,
            USA}}

\author[2]{Maziar Raissi}

\affil[2]{Department of Applied Mathematics,
            University of Colorado Boulder, 
            Boulder,
            610101, 
            Colorado,
            USA}
            
\author[1]{Arezoo M. Ardekani}

\begin{document}

\maketitle

\begin{abstract}
    Viscoelastic fluids are a class of fluids that exhibit both viscous and elastic nature. Modelling such fluids requires constitutive equations for the stress, and choosing the most appropriate constitutive relationship can be difficult. We present viscoelasticNet, a physics-informed deep learning framework that uses the velocity flow field to select the constitutive model and learn the stress field. Our framework requires data only for the velocity field, initial \& boundary conditions for the stress tensor, and the boundary condition for the pressure field. Using this information, we learn the model parameters, the pressure field, and the stress tensor. {This work considers} three commonly used non-linear viscoelastic models: Oldroyd-B, Giesekus, and linear  Phan-Thien-Tanner (PTT). We demonstrate that our framework works well with noisy and sparse data. Our framework can be combined with velocity fields acquired from experimental techniques like particle image velocimetry to get the pressure \& stress fields and model parameters for the constitutive equation. Once the model has been discovered using viscoelasticNet, the fluid can be simulated and modeled for further applications. 
\end{abstract}
\keywords{
Physics informed neural networks, Viscoelastic flow, Deep learning, Inverse modelling
}

\section{Introduction}

{Fluids can be categorized based on their response to the strain rate or the change in deformation with respect to time. For fluids that obey Newton's law of viscosity, the viscous stress at every point correlates linearly with the local strain rate. Numerous fluids, called non-Newtonian fluids, exhibit complex rheological behavior which deviates from Newton's law of viscosity. We can classify non-Newtonian fluids as inelastic, linear-viscoelastic, and non-linear viscoelastic fluids.} Viscoelastic fluids are a class of non-Newtonian fluids that exhibit viscous and elastic characteristics when subjected to deformation. These fluids are pertinent to various biological and industrial processes such as fertilization \cite{Li2021, Tung2017}, the collective motion of microorganisms \cite{Li2016, Li2014}, and oil recovery \cite{Hu2021, Wei2014}.

{The conservation of mass and momentum governs all fluid equations. The forces acting on the fluid are obtained for Newtonian fluids, assuming a linear correlation between stress and strain. However, viscoelastic fluids have both elastic and viscous characteristics.}  Hence, we need to solve a constitutive equation for stress along with the continuity and momentum equations. While linear viscoelastic models work well for small deformations, constitutive models that capture the non-linearity between stress and strain are required for large deformations. {These non-linear viscoelastic models can describe complex phenomena like shear thinning and extensional thickening.} Numerical methods like finite difference, finite elements, and finite volume are often {required} to obtain the stress field using these constitutive equations. However, non-linear viscoelastic models are often computationally demanding and require numerical tricks to ensure stability \cite{Alves2021, Beijer2002, Areias2008}. Moreover, selecting the most appropriate model for the fluid of interest can be challenging.

{Deep learning-based frameworks have helped solve challenging problems in various fields.} These include biomedical imaging \cite{Wang2016, Min2017}, computer vision \cite{Voulodimos2018, Esteva2021}, and natural language processing \cite{Young2018, Torfi2020}. There is growing interest in leveraging these techniques to understand and model biological and engineering systems. Machine learning algorithms have been used for problems in fluid mechanics for surrogate modeling, design optimization, and reduced order and closure models \cite{Brunton2020, Brunton2021, Wan2018, Fukami2020}. A deep neural network has been used to model viscoelastic properties from observed displacement data – as a PDE-constrained optimization challenge \cite{Xu2021}. However, many of these algorithms are data-intensive, and acquiring data at scale for engineering systems is often expensive.

{Physics-informed neural networks (PINNs) have emerged as a powerful tool in this context.} PINNs \cite{Raissi2019, DHP2018}, supervised learning frameworks with embedded physics, allow us to train massive neural networks with relatively small training datasets. PINNs achieve this data efficiency by using the governing equations to regularize the optimization of the neural network's parameters, enabling them to generalize even when few examples are available. While the most popular neural network architecture used for PINNs is a vanilla feed-forward neural network, {researchers have explored other architectures in the literature.} PINNs have been extended to use multiple feed-forward networks \cite{Haghighat2021, Moseley2021}, convolution neural networks \cite{Gao2021, Fang2021}, recurrent neural networks \cite{Zhang2020,Yucesan2021}, and Bayesian neural networks \cite{BPINNs2021}. {Researchers have used PINNs to help solve various forward and inverse problems in fluid mechanics} \cite{Jin2020, Arthurs2021, Cuomo2022}. Hidden fluid mechanics (HFM) \cite{Raissi2020}, a physics-informed deep learning framework, has been used to extract quantitative information from flow visualization. PINN-based frameworks have been used for solving Reynolds-averaged Navier Stokes equations \cite{Eivazi2022}, for modeling porous media flows \cite{Almajid2022}, and to solve inverse problems of three-dimensional supersonic and biomedical flows \cite{Cai2021}. Recently, a non-Newtonian PINNs-based framework was used for solving complex fluid systems \cite{Mahmoud2022}.

{Physics-informed neural networks (PINNs) can be extended along several dimensions.} These include: 1) more complex physics (i.e., equations), 2) more complex geometries, 3) better loss functions, 4) better architectures, and 5) better training processes. We are making contributions along dimensions 1, 3, and 5. {In this work, we present viscoelasticNet,} a physics-informed neural {networks-based} framework that uses the velocity flow field to select the viscoelastic constitutive model and learn the stress field. We consider three commonly used non-linear viscoelastic models: the Oldroyd-B \cite{OldroydB}, Giesekus \cite{Giesekus}, and Linear PTT \cite{PTT}. We combine the equations for these models into a single general equation. {We generate numerical data for each model mentioned above and employ our framework to learn the model parameters. Through this process, we showcase the capability of our framework to evaluate and select the most suitable model from the three considered models based on the learned parameters.} We also learn the pressure field and the stress tensor for the flow. The observables for our method are only the velocity field, the boundary and initial conditions for the stress field, and the boundary conditions for the pressure field. Hence, our method can be combined with experimentally acquired velocity fields to get the stress and pressure fields and select the viscoelastic constitutive equation for the fluid. {We discuss the problem setup and methodology in section \ref{sec:Methodology}.} We test our framework using the geometry of two-dimensional stenosis and a cross-slot geometry. We tested our framework for noise and sparsity in the velocity field using the stenosis geometry, and we used cross-slot geometry to carry out further tests on the effect of variation in parameters and the boundary conditions. Finally, we discuss the results in section \ref{sec:results} and provide some concluding remarks on our study in section \ref{sec:conclusion}.

\begin{figure}[!h]
    \centering
    \includegraphics[width=\linewidth]{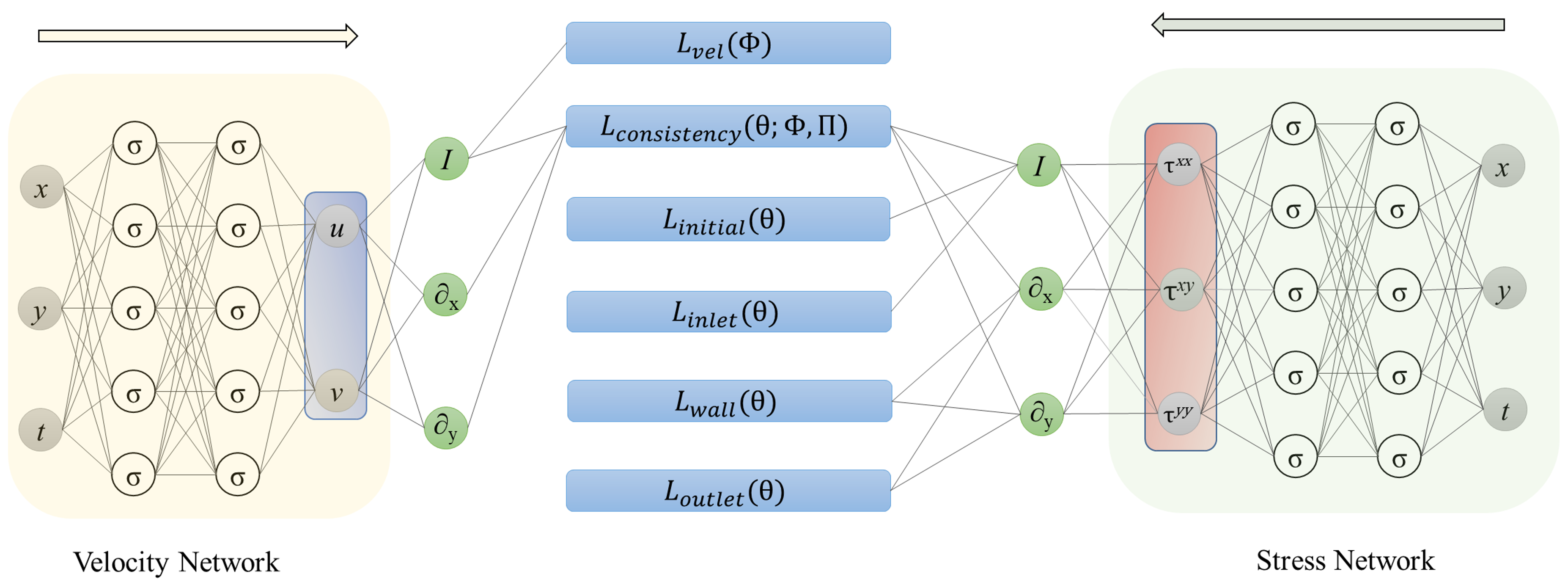}
    \caption{ {A schematic of the neural network set up to learn the stress field and parameters in eq. \eqref{genEq} by minimizing the loss function presented in eq. \eqref{stressLoss}. We use two fully connected neural networks to estimate the general constitutive equation's stress and parameters.} The network for velocity has an ivory color, while the network for stress has a green color, as shown in the figure. We use automatic differentiation to calculate the losses that we describe in section \ref{sec:Methodology}. We denote the identity operator by I and use automatic differentiation to compute the differential operators $\partial t,\partial x,\partial y$. }
    \label{fig:setup}
\end{figure}

\begin{figure}[!h]
    \centering
    
    \includegraphics[width=\linewidth]{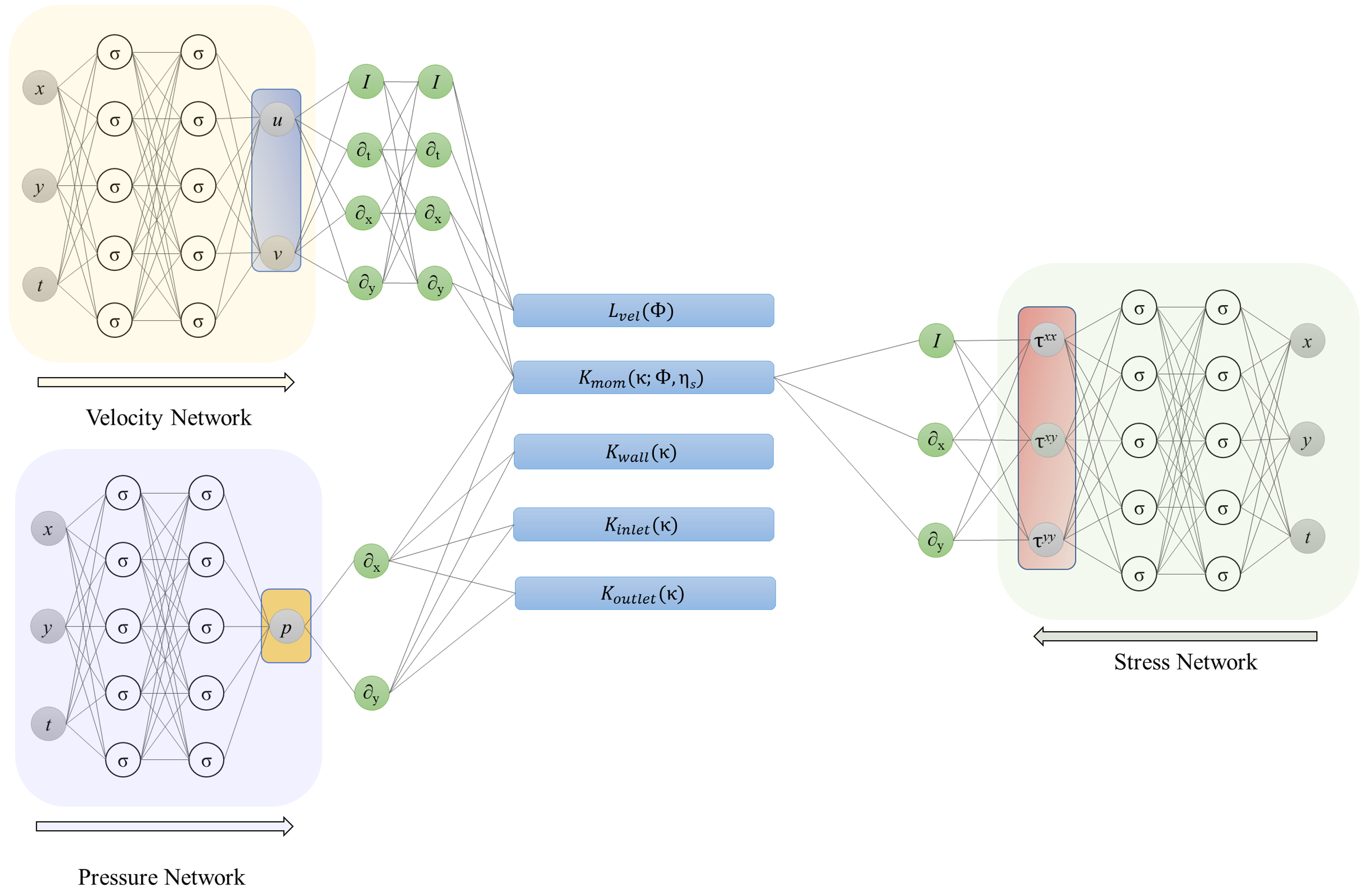}
    \caption{ {A schematic of the neural network set up to learn the pressure field and the viscosity by minimizing the loss function presented in eq. \eqref{pLoss}.} The figure shows three neural networks with different colors: ivory for velocity, green for stress, and purple for pressure. We fix the parameters of the stress network when we solve for the pressure. We calculate the losses using automatic differentiation, as we explain in section \ref{sec:Methodology}.   }
    \label{fig:setup}
\end{figure}
\section{Problem setup and methodology}
\label{sec:Methodology}

\subsection{Fluid motion equations}

Consider an incompressible fluid under isothermal, single-phase, transient conditions in a domain $\Omega \subset \mathbb{R}^d$ with boundary 
$\partial \Omega=\Gamma_D \cup \Gamma_N.$
The parameters $
\Gamma_D \text { and } \Gamma_N $ are portions of the boundary, respectively, where a Dirichlet and a Neumann boundary condition is applied, and d is the dimension. The following equations give the mass conservation and momentum balance in the absence of any body force

\begin{equation}\label{continuity}
    \nabla \cdot \boldsymbol{u} = 0,
\end{equation}

\begin{equation}\label{momentum}
    \rho \left( \frac{\partial \boldsymbol{u}}{\partial t} + \boldsymbol{u}\cdot \nabla \boldsymbol{u}\right) = -\nabla p + \nabla \cdot \boldsymbol{\tau'},
\end{equation}
where $\rho$ is the density of the fluid $\boldsymbol{u}$ is the velocity vector, $t$ is the time, $p$ is the pressure, and $\boldsymbol{\tau'}$ is the stress tensor. As for the boundaries, we have
\begin{equation}
\begin{cases}
\mathbf{u}=\mathbf{g} & \text { on } \Gamma_D \times(0, T) \\ 
\boldsymbol{\tau'}(\mathbf{u}, p) \hat{\mathbf{n}}=\mathbf{h} & \text { on } \Gamma_N \times(0, T) \\ \mathbf{u}(0)=\mathbf{u}_0 & \text { in } \Omega \times\{0\},\end{cases}
\end{equation}  
where $\hat{\mathbf{n}}$ is the outward directed unit normal, the functions $\mathbf{g}$ and $\mathbf{h}$ are given the Dirichlet and Neumann boundary data, respectively, and $\mathbf{u}_0$ is the initial condition. This work will represent scalars by non-bold characters, vectors by bold lowercase characters, and matrices by bold uppercase characters.

\subsection{Rheological constitutive model}
The stress tensor $\boldsymbol{\tau} '$ in eq. \eqref{momentum} for viscoelastic fluids is often split into solvent and polymeric parts,
\begin{equation}
    \boldsymbol{\tau'} = \boldsymbol{\tau}^s +  \boldsymbol{\tau}.
\end{equation}
We need a constitutive relation for the solvent and polymeric stress to have a well-posed problem. For a significant number of models, we can write the constitutive equations in the following form

\begin{equation}
    \boldsymbol{\tau}^s = \eta_s(\nabla \boldsymbol{u} + \nabla \boldsymbol{u}^T),
\end{equation}
\begin{equation}\label{conEq}
    f(\boldsymbol{\tau})\boldsymbol{\tau} + \lambda \overset{\nabla}{\boldsymbol{\tau}} + \boldsymbol{h}(\boldsymbol{\tau}) 
    = \eta_p (\nabla\boldsymbol{u} + \nabla \boldsymbol{u}^T),
\end{equation}
where we denote the solvent viscosity by $\eta_s$, the polymeric viscosity by $\eta_p$, the relaxation time by $\lambda$, the shear rate by $\dot \gamma$, $f(\boldsymbol{\tau})$ is a scalar-valued function, $\boldsymbol{h}(\boldsymbol{\tau})$ is a tensor-valued function and $\overset{\nabla}{\boldsymbol{\tau}}$ is the upper convected time derivative which is defined as 
\begin{equation}
    \overset{\nabla}{\boldsymbol{\tau}} = \frac{D\boldsymbol{\tau}}{Dt} - (\nabla\boldsymbol{u})^T \cdot \boldsymbol{\tau} - \boldsymbol{\tau} \cdot (\nabla\boldsymbol{u}),
\end{equation}
where
\begin{equation}
    \frac{D\boldsymbol{\tau}}{Dt} = \frac{\partial \boldsymbol{\tau}}{\partial t} + \boldsymbol{u}\cdot\nabla \boldsymbol{\tau}
\end{equation}
is the material derivative. The conservation of angular momentum principle implies that the polymetric stress tensor $\boldsymbol{\tau}$ is symmetric. Hence, we define the stress tensor $\boldsymbol{\tau}$ in two dimensions by three independent parameters
\begin{equation}
    \boldsymbol{\tau} = \begin{bmatrix}
    \tau^{xx} & \tau^{xy}  \\
    \tau^{xy}  & \tau^{yy} 
    \end{bmatrix},
\end{equation}
where $\tau^{xx} $ and $\tau^{yy} $ are the orthogonal normal stresses and $\tau^{xy} $ is the orthogonal shear stress. This work considers the Oldroyd-B \cite{OldroydB}, Giesekus \cite{Giesekus}, and Linear PTT \cite{PTT} models. The respective constitutive equations for the Oldroyd-B, Giesekus, and Linear PTT models are given by
\begin{equation}\label{OEq}
    \boldsymbol{\tau} + \lambda\overset{\nabla}{\boldsymbol{\tau}}
    = \eta_p(\nabla\boldsymbol{u} + \nabla \boldsymbol{u}^T),
\end{equation}
\begin{equation}\label{GEq}
    \boldsymbol{\tau} + \lambda\overset{\nabla}{\boldsymbol{\tau}} + \alpha \frac{\lambda}{\eta_p}(\boldsymbol{\tau} \cdot \boldsymbol{\tau})
    = \eta_p(\nabla\boldsymbol{u} + \nabla \boldsymbol{u}^T),
\end{equation}
and
\begin{equation}\label{PEq}
    \left(1+ \frac{\epsilon \lambda}{\eta_p}tr(\boldsymbol{\tau})\right)\boldsymbol{\tau} + \lambda\overset{\nabla}{\boldsymbol{\tau}}
    = \eta_p(\nabla\boldsymbol{u} + \nabla \boldsymbol{u}^T),
\end{equation}
where $tr(\boldsymbol{\tau})$ denotes the trace of the stress tensor, $\epsilon$ represents the extensibility parameter and $\alpha$ is the mobility parameter. We write the following general form equation   
\begin{equation}\label{genEq}
    \left(1+ \frac{\epsilon \lambda}{\eta_p}tr(\boldsymbol{\tau})\right)\boldsymbol{\tau} + \lambda\overset{\nabla}{\boldsymbol{\tau}}
    + \alpha \frac{\lambda}{\eta_p}(\boldsymbol{\tau} \cdot \boldsymbol{\tau})
    = \eta_p(\nabla\boldsymbol{u} + \nabla \boldsymbol{u}^T),
\end{equation}
which we use to represent the Oldroyd-B, Giesekus, and Linear PTT models. Equations \ref{OEq}, \ref{GEq}, \ref{PEq}, and \ref{genEq} are special cases of Eq. \ref{conEq}. As shown in table \ref{table:tabModels}, learning the values for the extensibility parameter ($\epsilon$) and the mobility parameter ($\alpha$) can help us select the constitutive equation {that best} describes the flow. {If $\epsilon$ and $\alpha$ equal zero, the Oldroyd-B model can describe the flow. }  Similarly, if $\epsilon$ or $\alpha$ are non-zero, the flow can be described using the linear PTT and Giesekus model, respectively. {If the learned values of both $\epsilon$ and $\alpha$ are non-zero, it implies that these three constitutive equations cannot describe the fluid. In this work, we demonstrate how the most appropriate model can be determined among these options based on the values of the learned parameters $\epsilon$ and $\alpha$.}
\begin{table}[ht]
  \caption{The list of parameters in eq. \eqref{genEq}  to represent the Oldroyd-B, Giesekus, and Linear PTT models.}
  \begin{center}
    \begin{tabular}{|c|c|c|}
      \hline
      \bf {Model} &\bf $\epsilon$ & \bf $\alpha$\\
      \hline
      Oldroyd & $0$ & $0$ \\
      \hline
      Gieseukus & $0$ & $\neq 0$ \\
      \hline
      Linear PTT & $\neq 0$ & $0$ \\
      \hline
    \end{tabular}
  \end{center}
  \label{table:tabModels}
\end{table}
\subsection{Physics informed neural networks}

We develop a physics-informed neural network-based framework called viscoelasticNet, which combines the information available in the velocity field, the Navier-Stokes equation, and the general form of the constitutive equation, eq. \eqref{genEq}. The objective is to learn the parameters of the constitutive equation while simultaneously solving the forward problem to obtain the stress field. We consider the velocity field $\boldsymbol{u}(t,\boldsymbol{x})=[u(t,\boldsymbol{x}), v(t,\boldsymbol{x})$] of an incompressible isothermal flow of a viscoelastic fluid, {where $\boldsymbol{x} = (x,y)$}. We observe N data points of time-space coordinates ($t^n, x^n, y^n$) and the velocity of the fluid corresponding to these points ($u^n, v^n$) where $n=1, \ldots, N$. Given such scattered spatiotemporal data, we are interested in the discovery of the components of the stress tensor $\boldsymbol{\tau}(t, \boldsymbol{x})$ as well as their governing equation by determining the parameters $\epsilon, \lambda, \alpha, \eta_p$ and $\eta_s$ in eq. \eqref{genEq}. Our setup has no input data on the pressure field and the stress tensor except for the initial and boundary conditions.
In our setup, we treat the $x$ and $y$ components of the velocity, the value of the stress field at the first time step (initial value), the stress field at the inlet, and the value of the pressure field at the outlet as the observable. We approximate the functions $(t, \boldsymbol{x}) \longmapsto (\sigma^{xx},\sigma^{xy},\sigma^{yy})$, $(t, \boldsymbol{x}) \longmapsto \psi$ and $(t, \boldsymbol{x}) \longmapsto p$ using three deep neural networks with parameters $\theta$, $\phi$ and $\kappa$ called the stress network, the velocity network and the pressure network, respectively. For the $x$-component of velocity  $u(t,\boldsymbol{x})$ and $y$-component $v(t,\boldsymbol{x})$, we define 
\begin{equation}
    u = \psi_y,    v = -\psi_x,
\end{equation}
for a scalar $\psi(t,\boldsymbol{x})$ and the subscripts represent partial derivatives. Defining the velocity field using a vector potential $\boldsymbol{\psi} = (0, 0, \psi)$ allows us to make the velocity field divergence free by construction, as we define $\boldsymbol{u} = \nabla \times \boldsymbol{\psi}$. This approach can be extended to three dimensions as well. The velocity field then automatically satisfies the continuity equation, eq. \eqref{continuity}. {We utilize a neural network to represent the velocity field because it enables us to compute derivatives of the velocity components with respect to the inputs, facilitating the calculation of residuals for the equations. While derivatives could be approximated using numerical methods like the finite difference approach, employing a neural network for the velocity field leverages automatic differentiation, which offers superior accuracy, efficiency, and stability compared to other numerical techniques.} We decouple the momentum equations from the constitutive equations for the polymeric stress and sequentially solve them. We chose a separate network for pressure as, in our experience, this setup works better with our decoupled sequential approach, and it is a fairly common technique employed in computational fluid dynamics to decouple pressure from the momentum equations. We define the mean squared error loss for regression over the velocity field as 
\begin{equation}
    L_{vel}(\phi) = \mathbb{E}_{(t,\boldsymbol{x},\boldsymbol{u})} \left[\frac{|\boldsymbol{u}(t,\boldsymbol{x};\phi) - \boldsymbol{u}|^2}{{\sigma_{\boldsymbol{u}}}^2} \right],
\end{equation}
where $\boldsymbol{u}$ is the reference velocity field, $\boldsymbol{u}(t,\boldsymbol{x};\phi)$ is the prediction from the network, and ${\sigma_{\boldsymbol{u}}}$ is the standard deviation of the reference velocity field, and $\mathbb{E}$ denotes the expectation approximated by the population mean (i.e., mean of the observations $t_n, x_n, y_n, u_n, v_n$  {where $n=1, \ldots, M$ for $M$ observations}). 
Since we are also solving the forward problem of learning the stress field, initial and boundary conditions on the stress field are required. We enforce the initial condition using the loss function
\begin{equation}
\begin{split}
    L_{init}(\theta) =  & \mathbb{E}_{(t^{init},\boldsymbol{x}^{init},{\boldsymbol{\tau}}^{init})} \left[\frac{|{\boldsymbol{\tau}}(t^{init},\boldsymbol{x}^{init};\theta)-{\boldsymbol{\tau}}^{init}|^2}{{\sigma_{\boldsymbol{\tau}}}^2} \right],
\end{split}
\end{equation}
where {$t^{init},\boldsymbol{x}^{init}$} is the spatio-temporal point cloud at the initial timestep, $\boldsymbol{\tau}^{init}$ is the stress field at the first time step $t^{init}$ and ${\sigma_{\boldsymbol{\tau}}}$ is standard deviation of $\boldsymbol{\tau}^{init}$. {We define $t^{init}$ as the initial timestep. It can be 0, or any other value that the user chooses which corresponds to the $\boldsymbol{\tau}^{init}$ being considered. For brevity, we define $\Pi = (\lambda,\epsilon,\alpha, \eta_p)$.} Now, let
\begin{equation}\label{f_val}
    \begin{split}
    \mathbf{f}(t, \boldsymbol{x}; \theta, \phi, \Pi) =  & \left(\frac{1}{\lambda}+ \frac{\epsilon}{\eta_p}tr(\boldsymbol{\tau})\right)\boldsymbol{\tau} + \boldsymbol{u}\cdot\nabla \boldsymbol{\tau} - (\nabla\boldsymbol{u})^T \cdot \boldsymbol{\tau} - \boldsymbol{\tau} \cdot (\nabla\boldsymbol{u})\\
    &+  \frac{\alpha}{\eta_p}(\boldsymbol{\tau} \cdot \boldsymbol{\tau})
    - \frac{\eta_p}{\lambda}(\nabla\boldsymbol{u} + \nabla \boldsymbol{u}^T).
    \end{split}
\end{equation}
{Eq. \eqref{f_val} represents the value of $\frac{\partial \boldsymbol{\tau}}{\partial t}$ in eq. \eqref{genEq}, and this definition allows us to use the backward Euler discretization to construct a ``physics-informed" network.} The output of the feed-forward networks will be called ``physics uninformed" in the rest of the text and denoted with a superscript ``pu".  We then create a physics-informed neural network using the backward Euler discretization

\begin{equation}
\begin{split}
    \boldsymbol{\tau}^{pi}(t, \boldsymbol{x}; \Delta t, \theta, \phi, \Pi) =  \boldsymbol{\tau}^{pu}(t+\Delta t, \boldsymbol{x};\theta) 
     + \Delta t  \mathbf{f}(t + &\Delta t, \boldsymbol{x}; \\
     & \theta, \phi, \Pi),
\end{split}
\end{equation}
where the superscript ``pi" is used to denote ``physics-informed". Since the physics-informed and uninformed networks evaluate the stress at the same point $(t,\boldsymbol{x})$, they {must} be consistent. We enforce this using a consistency loss
\begin{equation}
\begin{split}
    L_{consistency}(\theta;\Delta t, \phi, \Pi) =  \mathbb{E}_{(t, \boldsymbol{x})} \left[\frac{|{\boldsymbol{\tau}}^{pi}(t, \boldsymbol{x}; \Delta t, \theta, \phi, \Pi)-{\boldsymbol{\tau}}^{pu}(t, \boldsymbol{x};\theta)|^2}{{\sigma_{\boldsymbol{\tau}}}^2} \right],
\end{split}
\end{equation}
\begin{figure}[!h]
    \centering
    \includegraphics[width=\linewidth]{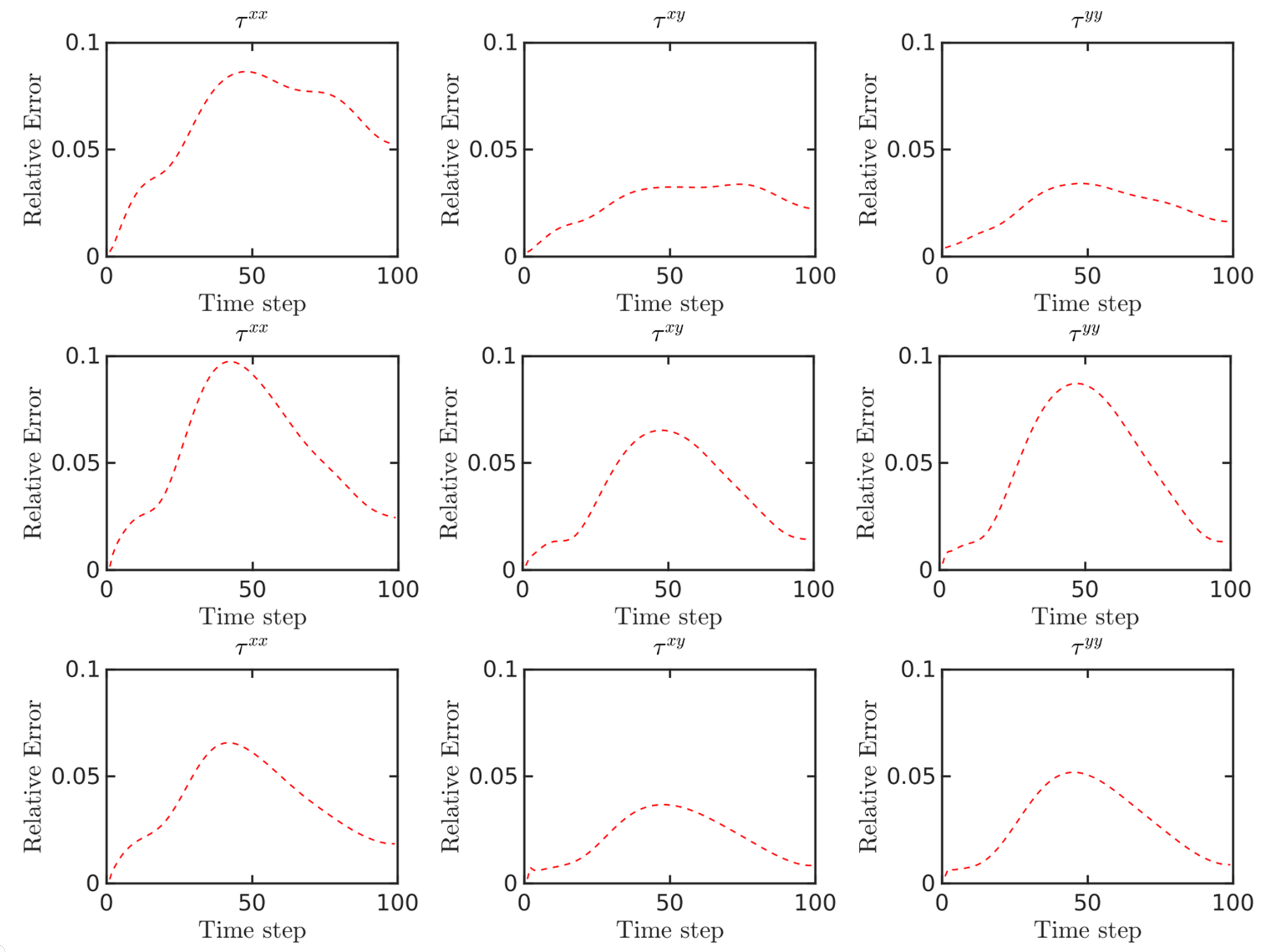}
    \caption{Relative errors between the predictions of the model and the corresponding reference components of the stress field across time steps for  (A) Oldroyd (B) Linear PTT (C) Giesekus constitutive models. }
    \label{fig:stressError}
\end{figure}
\begin{algorithm}[!h]
\caption{The algorithm for viscoelasticNet}\label{alg:one} 
   \textbf{Input:} Spatio-temporal point clouds, IC and BCs on stress
	\begin{algorithmic}[1]
            \State $\theta, \phi\leftarrow\theta^0, \phi^0$ \Comment{Initialize the neural network parameters}
		\For {$iteration=1,2,\ldots$} \Comment{Loop till the number of iterations}
                \State Compute $L_{stress}(\theta,\phi)$
                \State Update learning rate
                \State $\theta, \phi, \alpha, \epsilon, \lambda, \eta_p \leftarrow$ Optimizer($L_{stress}(\theta,\phi)$, learning rate) 
            \EndFor
            \State Freeze the optimized parameter $\theta '$
            \For {$iteration=1,2,\ldots$} \Comment{Loop till the number of iterations}
                \State Compute $K_{pressure}(\phi,\kappa)$
                \State Update learning rate
                \State $\phi,\kappa, \eta_s\leftarrow$ Optimizer($K_{pressure} (\phi,\kappa)$, learning rate) 
            \EndFor
	\end{algorithmic} 
    \textbf{Output:} $\theta', \phi', \kappa', \alpha', \epsilon', \lambda', \eta_p', \eta_s'$ \Comment{Optimized parameters}
\end{algorithm}
In this work, we utilize backward  Euler time-stepping to determine the relative weights for the loss terms based on the standard deviation of the available data. Using the standard deviation provides us with an equation-specific scale. We add the case-specific Neumann and Dirichlet boundary conditions for stress, {$L_{Neumann}(\theta)$ and $L_{Dirichlet}(\theta)$, respectively.} The parameters $\theta$ and $\phi$ are then optimized {along with $\lambda, \epsilon,$ and $\eta_p$} to minimize the following combined loss
\begin{equation}\label{stressLoss}
\begin{split}
    L_{stress}(\theta,\phi; \Pi) = & L_{vel}(\phi) + L_{initial}(\theta) + L_{Neumann}(\theta) +\\ 
    & L_{Dirichlet}(\theta) + L_{consistency}(\theta; \phi, \Pi).
\end{split}
\end{equation}
We show the schematic of the network in figure \ref{fig:setup}, and the algorithm for viscoelasticNet is explained using the pseudo algorithm \ref{alg:one}. Regularization is a practice used to avoid overfitting in machine learning. We use our prior knowledge of the governing equations to regularize the optimization process of the neural network parameters, as $L_{consistency}$ penalizes solutions that do not satisfy the governing equation. In this work, we utilize backward Euler time-stepping to determine the relative weights for the loss terms based on the standard deviation of the available data. We experimented with other techniques to obtain the loss weights, including assigning gradient-based weights and applying Lagrange multipliers. {However, we found that the backward Euler method performed best for our application.} Since we are sequentially solving the problem, we freeze the optimized parameters $\theta '$ of the neural network for the stress while solving for pressure. We split the momentum equation, eq. \eqref{momentum}, into two parts,  one which can be directly computed from the observables and the second which has unknown components. The convective part of the momentum equations is given by 
\begin{equation}
        \boldsymbol{g}^{L}(\boldsymbol{u}) =  \frac{\partial \boldsymbol{u}}{\partial t} + \boldsymbol{u}\cdot \nabla \boldsymbol{u},
\end{equation}
and
\begin{equation}
        \boldsymbol{g}^{R}(p;\boldsymbol{u},\boldsymbol{\tau},\eta_s) = -\nabla p  + \eta_s(\nabla^2 \boldsymbol{u} ) +  \nabla \cdot  \boldsymbol{\tau}.
\end{equation}
We calculate the standard deviation of $\boldsymbol{g}^{L}$ as ${\sigma_g}$. We then enforce the momentum equations, eq. \eqref{momentum}, using
\begin{equation}
    K_{m}(\kappa; \phi, \eta_s) =  \mathbb{E}_{(t,\boldsymbol{x})}\left[\dfrac{\left|\splitdfrac{\boldsymbol{g}^L(\boldsymbol{u}(t, \boldsymbol{x}; \phi))-\boldsymbol{g}^R( p(t, \boldsymbol{x}; \kappa);\boldsymbol{u}(t, \boldsymbol{x}; \phi),}{ \boldsymbol{\tau}(t,\boldsymbol{x};\theta '),\ \eta_s)}\right|^2}{\sigma_g^2}\right].
\end{equation}
We add the case-specific Neumman and Dirichlet boundary conditions for pressure {($K_{Neumann}(\kappa)$ and $K_{Dirichlet}(\kappa)$, respectively)} and optimize the parameters $\phi$ and $\kappa$ {along with $\eta_s$} using the following combined loss
\begin{equation}\label{pLoss}
    K_{pressure} (\phi, \kappa; \eta_s) = L_{vel}(\phi) + K_{mom}(\kappa; \phi, \eta_s) + K_{Neumann}(\kappa) + K_{Dirichlet}(\kappa).
\end{equation}

\begin{figure}
    \centering
    \includegraphics[width=\linewidth]{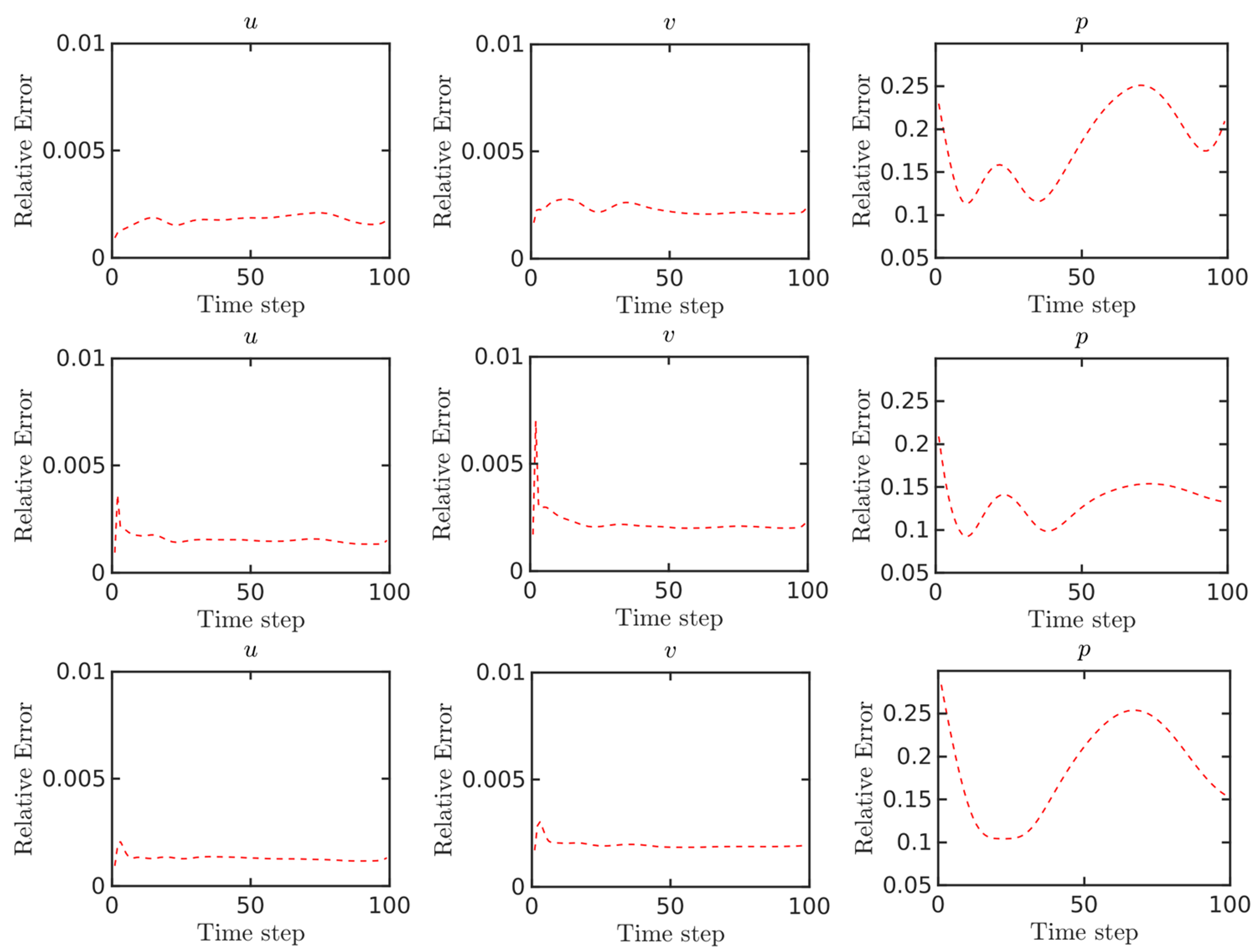}
    \caption{Relative errors between the predictions of the model and the corresponding reference velocity and pressure fields across the time steps for (A) Oldroyd (B) Linear PTT (C) Giesekus constitutive models.}
    \label{fig:uvpError}
\end{figure}
\begin{figure}
    \centering
    \includegraphics[width=8cm]{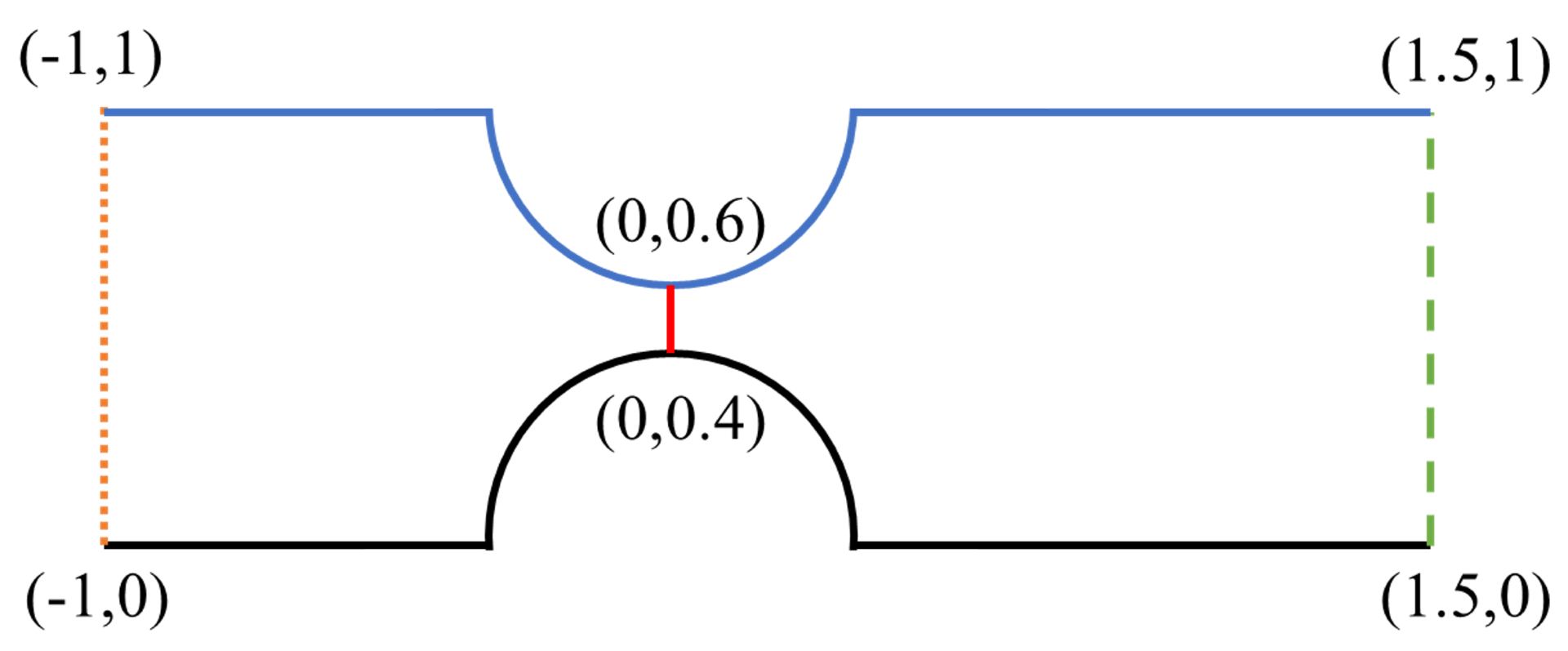}
    \caption{A two-dimensional stenosis. We show the domain walls using solid blue curves, the inlet with a dotted orange line, and the outlet with a dashed green line. {The lower wall of the stenosis is highlighted using a black line, and we plot the stress on the lower wall in Fig. \ref{fig:wall_stress}.} The narrowest part of the throat of the stenosis is highlighted with a red line. We plot the stress in this region in Fig. \ref{fig:stenosis_stress}.   }
    \label{fig:stenosis_throat}
\end{figure}

\section{Results}
\label{results}

\subsection{Stenosis}\label{stenosis}
We consider a two-dimensional {stenosis geometry}, as shown in Fig. \ref{fig:stenosis_throat}. We used RheoTool \cite{rheotool}, an OpenFOAM \cite{Weller198} based open source software developed by Favero et al. \cite{Favero2010}  to generate the training and reference data sets. RheoTool uses the finite volume method to discretize the equations. It uses the both-side-diffusion technique to increase the ellipticity, {stabilizing} the momentum equation. We use the log-confirmation tensor approach to tackle the numerical instabilities in the polymeric stress. More details on the solver and the validation for the code can be found here \cite{rheotool, Favero2010}. {We consider the stenosis in this work as the flow through stenotic vessels exhibits complex and interesting behavior. It is a challenging yet realistic scenario, as blood can be a viscoelastic fluid. Moreover, modeling flow in stenotic vessels can provide insights into hemodynamic parameters such as shear and wall stress, which can be clinically relevant.} The input to the algorithm essentially is the velocity field and the boundary conditions on stress and pressure.  
\begin{figure}
    \centering
    \includegraphics[width=\linewidth]{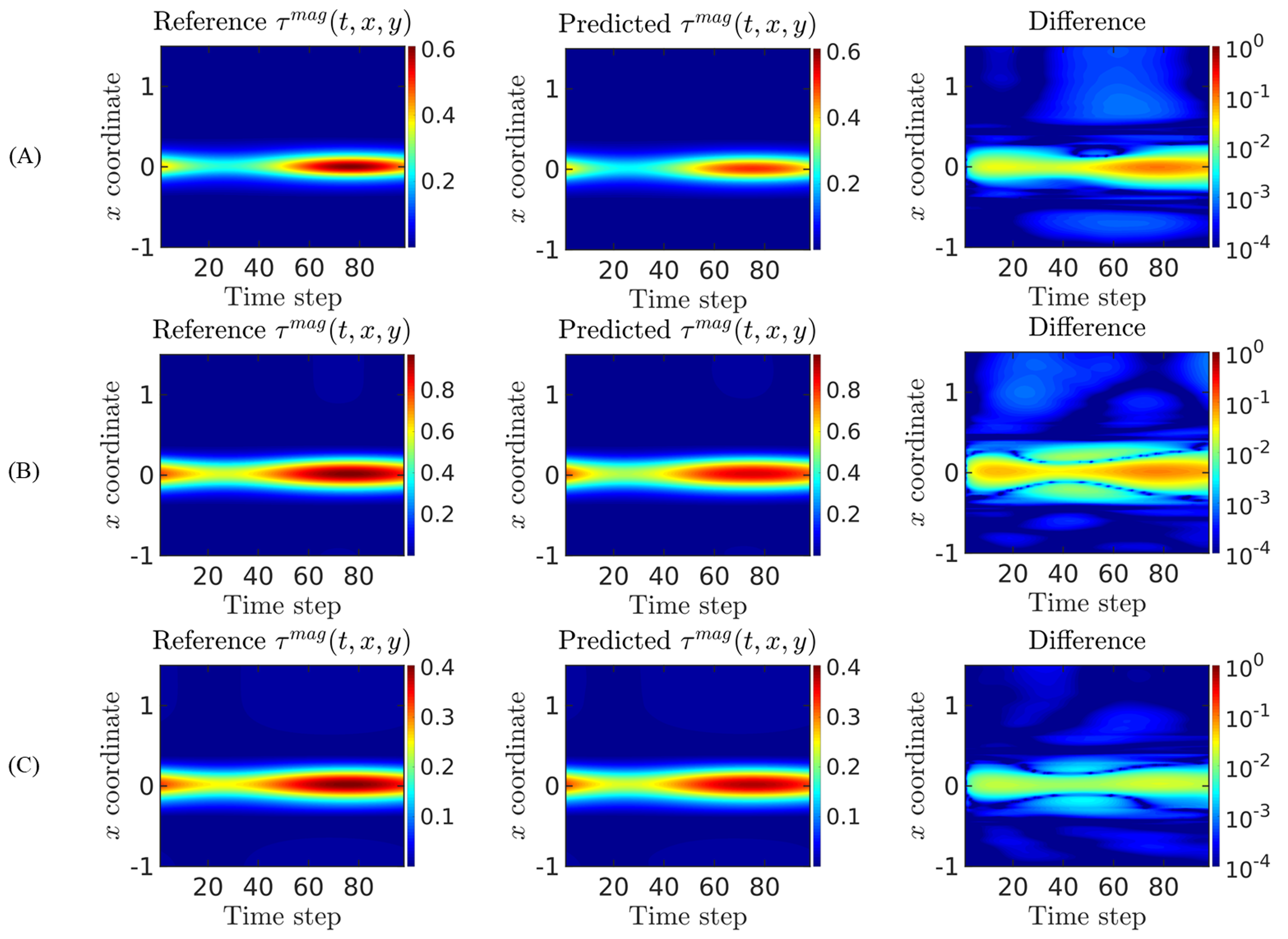}
    \caption{Comparison of the reference and predicted magnitude of the stress on the lower wall across all the time steps for (A) Oldroyd (B) Linear PTT (C) Giesekus constitutive models. The first column shows the reference values and the second column shows the model's predictions. {In the third column, we highlight the differences between the reference values and the model's predictions on a logarithmic scale.}}
    \label{fig:wall_stress}
\end{figure}
For all the results discussed in this section, we represent the {velocity components} ($u,v$) using an eight-layer deep, fully connected neural network with 128 neurons per hidden layer. We represent the stress components ($\tau^{xx}, \tau^{xy}, \tau^{yy}$) with another eight layers deep, fully connected neural network with 128 neurons per hidden layer. We use a third neural network to represent the pressure ($p$) with an eight-layer deep, fully connected neural network with 64 neurons per hidden layer. {We use fully connected neural networks with a constant number of neurons in each layer as, in our experience, such neural network architecture has worked better than increasing or decreasing neurons in each layer for PINNs.} All the networks use weight normalization but do not use batch normalization or dropout. We use the swish function as the activation function for all the networks. The swish activation function returns $x \times S(x)$ for an input $x$ and is known to match or outperform the ReLU activation function consistently.  {The sigmoid function ($S(x)$) is defined as $S(x) = \frac{1}{1+e^{-x}} $. The ReLU function is mathematically defined as ReLU(x) $=max(0,x)$. Future work could explore other architectures, such as convolutional neural networks, which may improve the results presented in this section. }

\begin{table}[ht]
  \caption{Relative error for flow variables at different noise levels for the Oldroyd-B model}
  \begin{center}
    \begin{tabular}{|c|c|c|c|}
      \hline
      \bf {} &\bf $u$&\bf $v$&\bf $p$\\
      \hline
      0\% Noise & $1.6\times10^{-3}$ & $2.2\times10^{-3}$& $1.94\times10^{-1}$\\
      \hline
      1\% Noise & $1.7\times10^{-3}$ & $2.3\times10^{-3}$& $2.02\times10^{-1}$\\
      \hline
      5\% Noise & $1.7\times10^{-3}$ & $2.4\times10^{-3}$& $1.87\times10^{-1}$\\
      \hline
      10\% Noise & $1.8\times10^{-3}$ & $2.6\times10^{-3}$& $1.69\times10^{-1}$\\
      \hline

    \end{tabular}
  \end{center}
  \label{table:Oldroyd_uvp}
\end{table}
\begin{table}[ht]
  \caption{Relative error for the stress components at different noise levels for the Oldroyd-B model}
  \begin{center}
    \begin{tabular}{|c|c|c|c|}
      \hline
      \bf {} &\bf $\tau^{xx}$& \bf $\tau^{xy}$& \bf $\tau^{yy}$\\
      \hline
      0\% Noise & $5.99\times10^{-2}$& $2.45\times10^{-2}$& $2.28\times10^{-2}$\\
      \hline
      1\% Noise & $5.93\times10^{-2}$ & $2.48\times10^{-2}$& $2.31\times10^{-2}$\\
      \hline
      5\% Noise & $5.94\times10^{-2}$ & $2.47\times10^{-2}$& $2.10\times10^{-2}$\\
      \hline
      10\% Noise & $6.02\times10^{-2}$ & $2.55\times10^{-2}$& $2.36\times10^{-2}$\\
      \hline

    \end{tabular}
  \end{center}
  \label{table:Oldroyd_stress}
\end{table}

\begin{table}[ht]
  \caption{Sensitivity to noise level in the velocity data for the Oldroyd model}
  \begin{center}
    \begin{tabular}{|c|c|c|c|c|c|}
      \hline
      \bf {} &\bf $\alpha$& \bf $\epsilon$& \bf $\lambda$& \bf $\eta_p$& \bf $\eta_s$\\
      \hline
      Reference value & $0.00$ & $0.0$& $0.05$& $0.008$& $0.01$\\
      \hline
      0\% noise & $0.00$ & $0.0$& $0.0517$& $0.0081$& $0.0098$\\
      \hline
      1\% noise & $0.00$ & $0.0$& $0.0517$& $0.0081$& $0.0115$\\
      \hline
      5\% noise & $0.00$ & $0.0$& $0.0517$& $0.0081$& $0.0115$\\
      \hline
      10\% noise & $0.00$ & $0.0$& $0.0517$& $0.0081$& $0.0112$\\
      \hline

    \end{tabular}
  \end{center}
  \label{table:Oldroyd_modelParams}
\end{table}

\begin{table}[ht]
  \caption{Relative error for flow variables at different noise levels for the linear PTT model}
  \begin{center}
    \begin{tabular}{|c|c|c|c|}
      \hline
      \bf {} &\bf $u$&\bf $v$&\bf $p$\\
      \hline
      0\% Noise & $1.6\times10^{-3}$ & $2.2\times10^{-3}$& $1.33\times10^{-1}$\\
      \hline
      1\% Noise & $1.8\times10^{-3}$ & $2.4\times10^{-3}$& $1.69\times10^{-1}$\\
      \hline
      5\% Noise & $1.9\times10^{-3}$ & $2.6\times10^{-3}$& $1.77\times10^{-1}$\\
      \hline
      10\% Noise & $2.2\times10^{-3}$ & $3\times10^{-3}$& $1.85\times10^{-1}$\\
      \hline

    \end{tabular}
  \end{center}
  \label{table:PTT_uvp}
\end{table}
\begin{table}[ht]
  \caption{Relative error for the stress components at different noise levels for the linear PTT model}
  \begin{center}
    \begin{tabular}{|c|c|c|c|}
      \hline
      \bf {} &\bf $\tau^{xx}$& \bf $\tau^{xy}$& \bf $\tau^{yy}$\\
      \hline
      0\% Noise & $5.45\times10^{-2}$& $3.64\times10^{-2}$& $4.62\times10^{-2}$\\
      \hline
      1\% Noise & $5.65\times10^{-2}$ & $3.77\times10^{-2}$& $4.7\times10^{-2}$\\
      \hline
      5\% Noise & $5.66\times10^{-2}$ & $3.82\times10^{-2}$& $4.75\times10^{-2}$\\
      \hline
      10\% Noise & $5.91\times10^{-2}$ & $3.97\times10^{-2}$& $4.78\times10^{-2}$\\
      \hline

    \end{tabular}
  \end{center}
  \label{table:PTT_stress}
\end{table}

\begin{table}[ht]
  \caption{Sensitivity to the noise level in the velocity data for the linear PTT model}
  \begin{center}
    \begin{tabular}{|c|c|c|c|c|c|}
      \hline
      \bf  &\bf $\alpha$& \bf $\epsilon$& \bf $\lambda$& \bf $\eta_p$& \bf $\eta_s$\\
      \hline
      Reference value & $0.00$ & $0.1$& $0.15$& $0.015$& $0.01$\\
      \hline
      0\% noise & $0.00$ & $0.106$& $0.161$& $0.0157$& $0.0097$\\
      \hline
      1\% noise & $0.00$ & $0.108$& $0.162$& $0.0157$& $0.0123$\\
      \hline
      5\% noise & $0.00$ & $0.108$& $0.162$& $0.0157$& $0.0123$\\
      \hline
      10\% noise & $0.00$ & $0.108$& $0.161$& $0.0157$& $0.0127$\\
      \hline

    \end{tabular}
  \end{center}
  \label{table:PTT_model_params}
\end{table}

\begin{table}[ht]
  \caption{Relative error for flow variables at different noise levels for the Giesekus model}
  \begin{center}
    \begin{tabular}{|c|c|c|c|}
      \hline
      \bf {} &\bf $u$&\bf $v$&\bf $p$\\
      \hline
      0\% Noise & $1.3\times10^{-3}$ & $2.0\times10^{-3}$& $1.95\times10^{-1}$\\
      \hline
      1\% Noise & $1.4\times10^{-3}$ & $2.1\times10^{-3}$& $1.82\times10^{-1}$\\
      \hline
      5\% Noise & $1.4\times10^{-3}$ & $2.2\times10^{-3}$& $1.83\times10^{-1}$\\
      \hline
      10\% Noise & $1.7\times10^{-3}$ & $2.5\times10^{-3}$& $1.77\times10^{-1}$\\
      \hline

    \end{tabular}
  \end{center}
  \label{table:Giesekus_uvp}
\end{table}

\begin{figure}
    \centering
    \includegraphics[width=\linewidth]{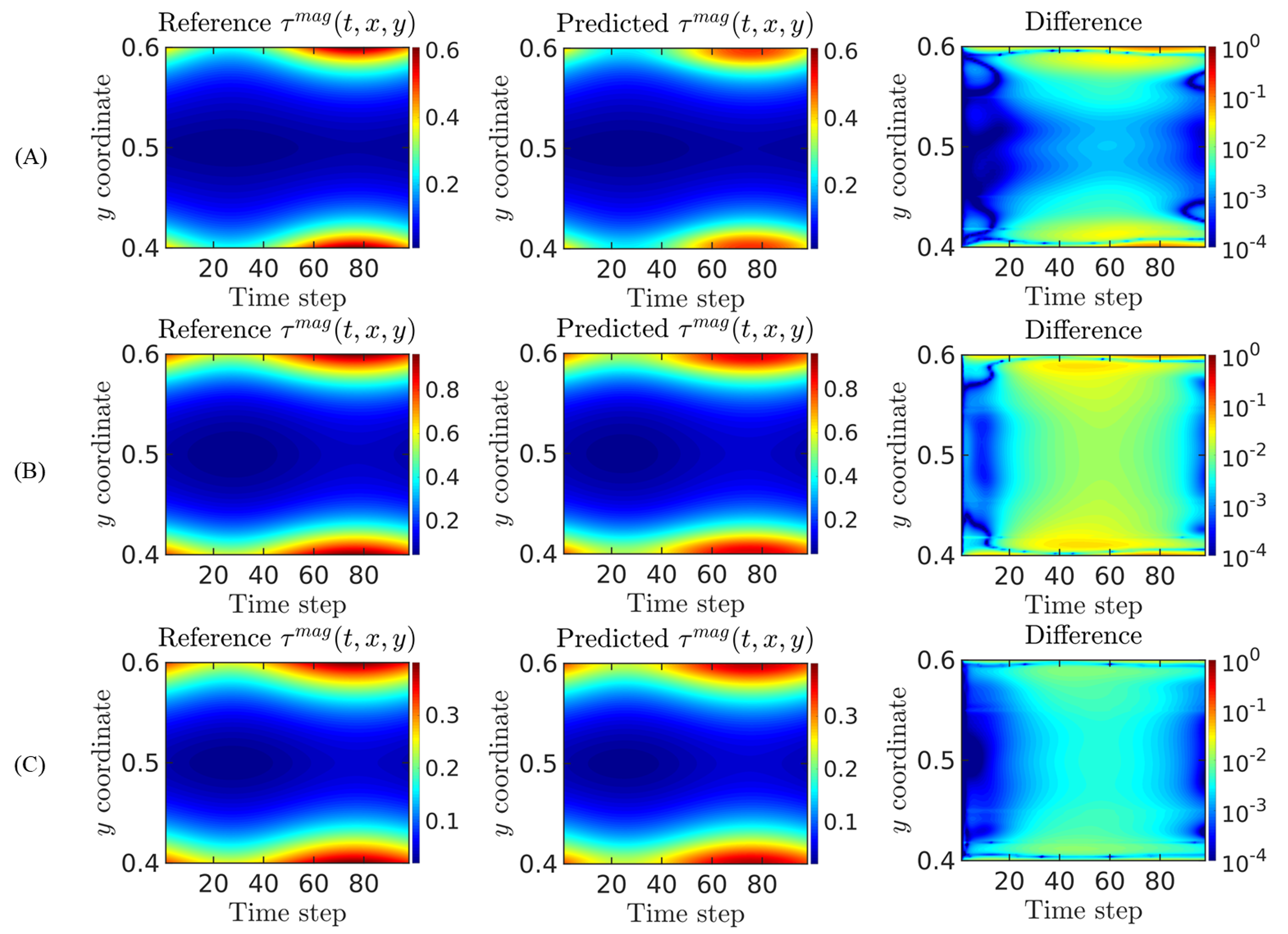}
    \caption{ Comparison of the reference and predicted magnitude of the stress in the throat of the stenosis across all the timesteps for (A) Oldroyd (B) Linear PTT (C) Giesekus constitutive models. The first column shows the reference values, and the second column shows the model's predictions. {In the third column, we highlight the differences between the reference values and the model's predictions on a logarithmic scale.}}
    \label{fig:stenosis_stress}
\end{figure}
The learning rate schedule is an important hyperparameter that determines how well the network parameters are optimized. For all the results reported in this work, we use a cosine annealing learning rate schedule \cite{Ilya2017}. The annealing learning rate schedule starts with a large learning rate{, gradually decreasing} to the defined minimum value. This allows for exploration while optimizing the parameters, and the reduction in the learning rate value refines the search close to the optima. We used a value of $2.5\text{e-}03$ for $\zeta_{max}$ and $2.5\text{e-}06$ for $\zeta_{min}$ to get the learning rate $\zeta$ as defined in the following equation
\begin{equation}\label{LR}
    \zeta =  \zeta_{min} + 0.5(\zeta_{max} - \zeta_{min})\left( 1 + cos\left( \frac{T_{cur}}{T_{max}}\pi\right)\right),
\end{equation}
where $T_{cur}$ is the current time step and $T_{max}$ is the total time step.
For learning the parameters in the general equation, eq. \eqref{genEq}, and the stress field, two million iterations of the Adam optimizer \cite{Kingma2015} were used. As we sequentially solve for the pressure, we first optimize the parameters of the neural network for the stress ($\theta$) by minimizing the loss function defined in eq. \eqref{stressLoss} and freezing them. We then optimize the parameters for the neural network for velocity ($\phi$) and pressure ($\kappa$) by minimizing the loss specified in eq. \eqref{pLoss}. {We ran 800,000 iterations of the Adam optimizer} for this optimization process with the same learning rate schedule defined above. We investigated different learning rate schedules, such as different values for the maximum and minimum values for the learning rate, while using cosine annealing and a step function to decay the learning rate. It was observed that using the same learning rate for both trainings leads to better results. A description of the loss function used to train the model is shared in \ref{sec:sample:appendix}.
We consider the geometry of a 2D stenosis as shown in figure \ref{fig:stenosis_throat} for all the results discussed in this section. While generating the reference dataset, we applied a sinusoidal boundary condition for the inlet velocity. The simulation ran for hundred time steps, or half a sine wave ( 0 to $\pi$). As discussed in section \ref{sec:Methodology}, we choose a sequential approach to solve for the stress and the pressure. Given the initial and boundary conditions on the stress, we are simultaneously solving the inverse problem of learning the parameters of the general equation, eq. \eqref{genEq}, and the forward problem of discovering the stress field in the spatio-temporal domain.  To compare the results predicted by the neural networks to the reference value and simulation results, we define the relative error to be

\begin{equation}\label{error}
    \mathcal{L}(a_{reference},a_{prediction}) = \sqrt{\frac{\overline{(a_{reference}-a_{prediction})^2}}{\overline{(a_{reference}-\Bar{a}_{reference})^2}}},
\end{equation}
where the bar denotes the mean value, we use this definition for error so that the multiplication or addition of a constant does not change it. We show the relative error between the predicted and reference values for the stress, velocity, and pressure fields in Fig. \ref{fig:stressError} and Fig. \ref{fig:uvpError}. As expected, the lowest errors are for the velocity fields, as there is data on those fields.  {The errors are lowest at the initial time steps since the initial condition for the stress is known. The non-monotonic nature of the errors is due to the sinusoidal boundary condition.} The agreement between the reference and predicted values is satisfactory as the mean relative error in the stress magnitude is less than 5\% for all cases. To test the effect of noise in boundary conditions on the model, we added 1\%, 5\%, 10\%, and 25\% Gaussian noise to the Dirichlet boundary condition on stress. Despite intentionally corrupting the boundary data, the model demonstrated resilience by learning an equivalent set of parameters across all noise levels. This insensitivity suggests that the model successfully captures the underlying dynamics instead of overfitting to specifics or noise in the boundary conditions. Since we solve for the pressure field in a decoupled manner, {errors in the stress field propagate, resulting in increased errors in the pressure field. Similar observations have been noted in other studies} \cite{Raissi2020}.

In Fig. \ref{fig:stenosis_stress}, we plot the reference and predicted values of the magnitude of the stress in the throat of the stenosis across all the time steps. Although there is an excellent qualitative and quantitative agreement between the predicted and reference values, the model seems to under-predict the magnitude of the stress on the walls. To focus on the stress on the walls, in Fig. \ref{fig:wall_stress}, we plot the reference and predicted values of the magnitude of stress on the lower wall across all the time steps for the Oldroyd, Linear PTT and Giesekus models. The stress magnitude on the lower and upper walls is symmetric, so we show the results only for the lower wall. The predictions for the Giesekus model perform best, with excellent qualitative and quantitative agreement between the reference and predicted values. However, the model under-predicts the peak value in all cases. 
 
{To check the robustness of our framework}, we add Gaussian noise to the velocity observations. The effect of noise on the parameters for the Oldroyd-B, linear PTT, and Giesekus models are reported in the tables \ref{table:Oldroyd_modelParams}, \ref{table:PTT_model_params}, and \ref{table:Giesekus_modelParams}, respectively. Adding Gaussian noise does not significantly affect the parameters learned for eq. \eqref{genEq}. However, there is an increase in the error for the learned viscosity $\eta_s$. Interestingly, the error does not increase as we increase the amount of Gaussian noise. {The reported values of $\epsilon$ and $\alpha$ illustrate how our framework facilitates model selection. All the learned values align consistently with the conditions specified in Table \ref{table:tabModels}. For the Oldroyd-B model, the learned values for both $\epsilon$ and $\alpha$ are equal to zero. In comparison, only the value for $\epsilon$ is zero for the Gieskus model, and only $\alpha$ equals zero for the Linear PTT models. If both the learned values of $\epsilon$ and $\alpha$ are nonzero, it implies that none of the three constitutive equations can model the fluid, and new constitutive equations need to be considered.} The error for the learned velocity, pressure, and stress components for the Oldroyd-B, linear PTT, and Giesekus models are reported in Tables \ref{table:Oldroyd_uvp}, \ref{table:Oldroyd_stress}, \ref{table:PTT_uvp}, \ref{table:PTT_stress}, \ref{table:Giesekus_uvp}, and \ref{table:Giesekus_stress}. The general trend is that the error for each variable increases slightly as the noise level increases, but the increase in error is not significant.

{We believe this low sensitivity to Gaussian noise occurs because the model uses many data points.} The models were trained on 5.78 million spatio-temporal data points of velocity. We tried training our model on fewer data points to test this hypothesis. Specifically, we consider the Giesekus model with 5.78 million, 578 thousand, 57.8 thousand, and 5.78 thousand data points with 5\% Gaussian noise in the velocity data. The results for the parameters are summarized in Table \ref{table:spatioTempData}. The results start to deteriorate at about 57.8 thousand points, with the value for viscosity ($\eta_s$) being off by about 50\%. The model fails to learn the viscosity ($\eta_s$) with 5.78 thousand spatiotemporal points but still learns the parameters of the general equation, eq. \ref{genEq}, reasonably well. We conducted this study to test the feasibility of using our framework with flow visualization techniques such as PIV. While the resolution can vary, about 500 spatial locations per time step is a realistic estimate of the resolution for a PIV experiment. Considering 500 spatial locations over 100 time steps, a realistic number would be getting 50,000 spatiotemporal points from an experiment.

{These results lead us to a promising conclusion that the model performs well with noisy and sparse datasets, a significant advantage considering the often noisy nature of experimental data and the challenges of acquiring high-resolution data. This opens up exciting possibilities for integrating our approach with experimentally acquired datasets. If the velocity field is obtained experimentally, our method can potentially learn the stress field and pressure field and select the appropriate constitutive equation among the discussed models, provided that the boundary conditions are known. This exciting capability paves the way for practical applications in experimental fluid mechanics and constitutive modeling. }

\begin{table}[ht]
  \caption{Relative error for the stress components at different noise levels for the Giesekus model}
  \begin{center}
    \begin{tabular}{|c|c|c|c|}
      \hline
      \bf {} &\bf $\tau^{xx}$& \bf $\tau^{xy}$& \bf $\tau^{yy}$\\
      \hline
      0\% Noise & $3.88\times10^{-2}$& $2.16\times10^{-2}$& $2.88\times10^{-2}$\\
      \hline
      1\% Noise & $3.75\times10^{-2}$ & $2.12\times10^{-2}$& $2.83\times10^{-2}$\\
      \hline
      5\% Noise & $3.83\times10^{-2}$ & $2.17\times10^{-2}$& $2.89\times10^{-2}$\\
      \hline
      10\% Noise & $3.85\times10^{-2}$ & $2.19\times10^{-2}$& $2.89\times10^{-2}$\\
      \hline

    \end{tabular}
  \end{center}
  \label{table:Giesekus_stress}
\end{table}

\begin{table}[ht]
  \caption{Sensitivity to the noise level in the velocity data for the Gieseukus model}
  \begin{center}
    \begin{tabular}{|c|c|c|c|c|c|}
      \hline
      \bf {} &\bf $\alpha$& \bf $\epsilon$& \bf $\lambda$& \bf $\eta_p$& \bf $\eta_s$\\
      \hline
      Reference value & $0.2$ & $0.0$& $0.1$& $0.01$& $0.01$\\
      \hline
      0\% noise & $0.205 $ & $0.0$& $0.105$& $0.0094$& $0.0103$\\
      \hline
      1\% noise & $0.205$ & $0.0$& $0.105$& $0.0098$& $0.0103$\\
      \hline
      5\% noise & $0.205$ & $0.0$& $0.105$& $0.0097$& $0.0103$\\
      \hline
      10\% noise & $0.205$ & $0.0$& $0.105$& $0.0099$& $0.0103$\\
      \hline

    \end{tabular}
  \end{center}
  \label{table:Giesekus_modelParams}
\end{table}

\begin{table}[H]
  \caption{Sensitivity to amount of spatio-temporal data}
  \begin{center}
    \begin{tabular}{|c|c|c|c|c|c|}
      \hline
      \bf {} &\bf $\alpha$& \bf $\epsilon$& \bf $\lambda$& \bf $\eta_p$& \bf $\eta_s$\\
      \hline
      Reference value & $0.2$ & $0.0$& $0.1$& $0.01$& $0.01$\\
      \hline
      5.78 million & $0.205$ & $0.0$& $0.105$& $0.0097$& $0.0103$\\
      \hline
      578 thousand & $0.205$ & $0.0$& $0.105$& $0.0103$& $0.0093$\\
      \hline
      57.8 thousand & $0.206$ & $0.0$& $0.106$& $0.0103$& $0.0155$\\
      \hline
      5.78 thousand & $0.217$ & $0.0$& $0.111$& $0.0104$& $0.00045$\\
      \hline

    \end{tabular}
  \end{center}
  \label{table:spatioTempData}
\end{table}

\subsection{Cross-slot}
\begin{figure}
    \centering
    \includegraphics[width=\linewidth]{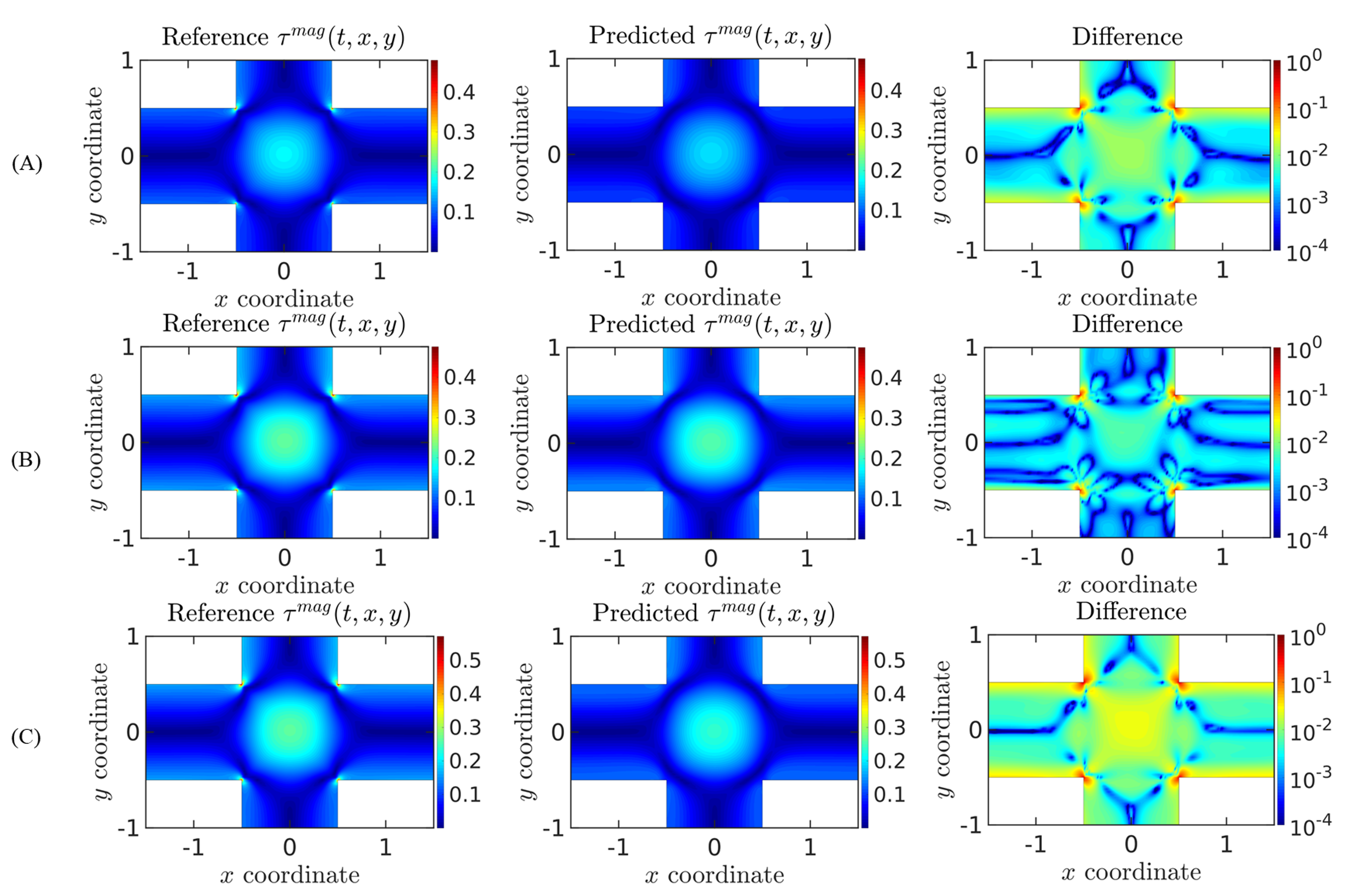}
    \caption{Comparison of the reference and predicted magnitude of the stress in the cross-slot at the 50th times step for (A) Oldroyd (B) Linear PTT (C) Giesekus constitutive models. The first column shows the reference values, the second column shows the model's predictions, and the third column shows the difference in the two values {on a logarithmic scale}.}
    \label{fig:crossSlot_stress}
\end{figure}
We now examine a cross-slot geometry, a popular test case for constitutive models of non-Newtonian fluids. We used RheoTool \cite{rheotool} to generate the reference dataset for this problem. As with the previous geometry, the velocity boundary condition at the inlets is transient and varies sinusoidally. The loss function used to train the model is shared in \ref{sec:sample:appendix}. We consider a hundred time steps as our reference dataset. The neural network architecture and input features were the same as those used for the stenosis problem in section \ref{stenosis}, and we again chose a sequential approach to solve for the stress and then for the pressure. We need two different learning rate schedules for this geometry. We used the cosine annealing learning rate described in eq. \eqref{LR}, but we used different values of $\zeta_{max}$ for the parameters $\alpha, \epsilon, \lambda, \eta_p$ and $\eta_s$ than for the weights and biases. The value of $\zeta_{max}$ for the parameters mentioned above was 2.5e-04, while it was 2.5e-3 for the weights and biases. The $\zeta_{min}$ value was 2.5e-06 for all the parameters. The output features were the velocity field, the stress field, and the pressure field. The boundary conditions for stress and pressure are the same as defined in section \ref{stenosis}, and the loss functions to solve for the stress and the pressure sequentially remain eqs. \eqref{stressLoss_stenosis} and \eqref{pLoss_stenosis}, respectively.
\begin{table}[ht]
  \caption{Relative error averaged across all time steps for the stress components for the cross-slot geometry for the Oldroyd-B, Linear PTT and Giesekus models.}
  \begin{center}
    \begin{tabular}{|c|c|c|c|}
      \hline
      \bf {} &\bf $\tau^{xx}$& \bf $\tau^{xy}$& \bf $\tau^{yy}$\\
      \hline
      Oldroyd-B& $1.80\times10^{-1}$ & $2.31\times10^{-1}$& $1.94\times10^{-1}$\\
      \hline
      Linear PTT & $1.29\times10^{-1}$ & $1.66\times10^{-1}$& $1.58\times10^{-1}$\\
      \hline
      Giesekus & $1.94\times10^{-1}$ & $2.37\times10^{-1}$& $2.01\times10^{-1}$\\
      \hline
    \end{tabular}
  \end{center}
  \label{table:crossSlot_tau}
\end{table}
\begin{table}[ht]
  \caption{Relative error averaged across all time steps for flow variables for the cross-slot geometry for the Oldroyd-B, Linear PTT and Giesekus models.}
  \begin{center}
    \begin{tabular}{|c|c|c|c|}
      \hline
      \bf {} &\bf $u$&\bf $v$&\bf $p$\\
      \hline
      Oldroyd-B & $2.46\times10^{-2}$ & $2.32\times10^{-2}$& $1.17\times10^{1}$\\
      \hline
      Linear PTT & $1.39\times10^{-2}$ & $1.23\times10^{-2}$& $1.17\times10^{1}$\\
      \hline
      Giesekus & $2.29\times10^{-2}$ & $2.24\times10^{-2}$& $1.19\times10^{1}$\\
      \hline
    \end{tabular}
  \end{center}
  \label{table:crossSlot_uvp}
\end{table}

Tables \ref{table:crossSlot_tau} and \ref{table:crossSlot_uvp} show the errors for the stress field and the flow variables, respectively. The errors have been computed using the description in eq. \eqref{error}. Fig. \ref{fig:crossSlot_stress} compares the reference and predicted stress magnitudes at the 50th time step.
Our model was able to estimate the parameters of eq. \eqref{genEq} with reasonable accuracy, but it had higher errors for the stress field than in the stenosis case. {The error in the stress field also affected the accuracy of the pressure field,} which depends on the stress field. However, our model captured the viscosity very well, as it accurately reproduces the stress field in most of the domain. The primary source of error was at the corners of the cross-slot, where the stress field had sharp peaks that our model could not capture. {This error happened because our model used a single global network}, which tended to smooth over these discontinuities, and we could not capture the peak value of stress at the corners. A possible way to overcome this limitation is to use multiple networks or domain decomposition, which can be explored in future work.
\begin{table}[!h]
  \caption{The reference and predicted values of the parameters for the  Gieseukus, linear PTT, and Oldroyd-B models for the cross-slot geometry. {The second dataset (\#2) of the linear PTT model considers doubling the flow rate while keeping the same parameters. To evaluate the effect of different parameter combinations on the efficacy of the framework, we examined three cases of the Oldroyd-B model. Additionally, we tested the framework's capability of model selection by applying it to the flow field obtained from an extended Pom-Pom (XPomPom) constitutive equation. }}
  \begin{center}
    \begin{tabular}{|c|c|c|c|c|c|}
      \hline
      \bf {} &\bf $\alpha$& \bf $\epsilon$& \bf $\lambda$& \bf $\eta_p$& \bf $\eta_s$\\
      \hline
      Giesekus Reference value & $0.05$ & $0.0$& $0.004$& $0.003$& $0.01$\\
      \hline
      Giesekus Predicted value & $0.056$ & $0.0$& $0.00386$& $0.0273$& $0.011$\\
      \hline
      Linear PTT Reference value & $0.0$ & $0.02$& $0.008$& $0.025$& $0.01$\\
      \hline
      Linear PTT Predicted value & $0.0$ & $0.0183$& $0.0085$& $0.0245$& $0.0099$\\
      \hline
      Linear PTT Reference value \#2& $0.0$ & $0.02$& $0.008$& $0.025$& $0.01$\\
      \hline
      Linear PTT Predicted value \#2& $0.0$ & $0.0228$& $ 0.00836 $& $ 0.0239 $& $ 0.0102$\\
      \hline
      Oldroyd-B Reference value & $0.0$ & $0.0$& $0.005$& $0.01$& $0.01$\\
      \hline
      Oldroyd-B Predicted value & $0.0$ & $0.0$& $0.0046$& $0.0188$& $0.011$\\
      \hline
      Oldroyd-B Reference value \#2& $0.0$ & $0.0$& $0.015$& $0.01$& $0.02$\\
      \hline
      Oldroyd-B Predicted value \#2& $0.0$ & $0.0$& $ 0.0135 $& $ 0.0193 $& $ 0.0236$\\
      \hline
      Oldroyd-B Reference value \#3& $0.0$ & $0.0$& $0.01$& $0.025$& $0.02$\\
      \hline
      Oldroyd-B Predicted value \#3& $0.0$ & $0.0$& $ 0.0093$& $ 0.033 $& $ 0.0171$\\
      \hline
      XPomPom Predicted value & $37.29$ & $0.00423$& $0.29$& $0.145$& $0.011$\\
      \hline
    \end{tabular}
  \end{center}
  \label{table:crossSlot}
\end{table}
Table \ref{table:crossSlot} shows the learned parameters of eq. \eqref{genEq}. {To test the sensitivity of our model to boundary conditions and strain rates, we doubled the flow rate for the second dataset of Linear PTT (\#2). It was observed that this increase did not significantly affect the values of the learned parameters. Our framework can be applied to linear and nonlinear regimes if the fluid follows one of the constitutive equations presented in this work. To evaluate the effectiveness of our framework in learning different parameter combinations, we tested it on three distinct parameter sets for the Oldroyd-B model. The framework was able to learn all three different sets of parameters and select the model accurately. Additionally, to assess our framework's capability for model selection, we tested it using velocity data from the extended Pom-Pom or XPomPom model \cite{Verbeeten2001}, which is not included among the three models considered in our study. The learned parameters did not satisfy any of the criteria listed in Table \ref{table:tabModels} as both $\epsilon$ and $\alpha$ had a non-zero value, indicating that none of the three models (Gieseukus, linear PTT, or Oldroyd-B models) was a suitable fit for this flow.} 

{In this study, we have considered three models and developed a framework to identify which of these three models best fits the data. This represents an important advancement in integrating machine learning and physics-informed neural networks to address challenges in the constitutive modeling of viscoelastic fluids. We have not expanded this work beyond these three constitutive equations since our forward solver has been developed only for these three constitutive equations. If none of these constitutive equations are appropriate for a dataset, the neural network will notify the user by the learned values of $\epsilon$ and $\alpha$. We encourage further developments based on the ideas presented in this paper to include a wider range of constitutive equations.  }

\section{Conclusions and future scope of work}
\label{conclusion}
Machine learning algorithms are proving to be an increasingly useful tool in solving problems in fluid mechanics. However, the cost of high-fidelity data often makes utilizing these data-intensive tools {impractical. We introduce viscoelasticNet, a physics-informed neural networks (PINNs)-based framework to address this. This framework selects the viscoelastic constitutive model and learns the stress field from a velocity flow field. We} work with three commonly used non-linear viscoelastic models: the Oldroyd-B, Giesekus, and Linear PTT, and build a generalized framework to model them. The velocity, pressure, and stress fields are represented using neural networks. The backward Euler method was used to construct PINNs for the viscoelastic constitutive model. We use a multistage approach to solve the problem by first solving for the stress and then using the stress and velocity fields to solve for the pressure. To test our framework, we used noisy and sparse data sets in this work. We observed that the framework could learn the parameters of the viscoelastic constitutive model reasonably well in all the cases. 

In this work, we {applied the viscoelasticNet framework to} a stenosis geometry in two dimensions with the above-mentioned constitutive models and also {examined} the flow in a cross-slot. While our framework could learn the constitutive equation parameters with reasonable accuracy for both cases, it did not capture the peak stress at the corners of the cross-slot well. {To address this, we propose exploring a smaller domain instead of a global function.} This framework {has the potential to} be extended to include other rheological constitutive models like the FENE-P and extended Pom-Pom models. {We also suggest learning the entire equation instead of just the parameters in a fixed constitutive equation. Future research could consider} more complex geometries and three-dimensional cases. The framework {we present} here can augment techniques like particle image velocimetry (PIV). While PIV can acquire the velocity flow field, {our method can acquire} the pressure and stress fields. Once the constitutive equation is learned, the parameters can be used to model any future applications of the pertinent fluid.

\section{Acknowledgements}

A.M.A. acknowledges financial support from the National Science Foundation (NSF) through  Grant No.  CBET-2141404.

\section{Details on training}
\label{sec:sample:appendix}

\begin{figure}[!h]
    \centering 
    \includegraphics[width=\linewidth]{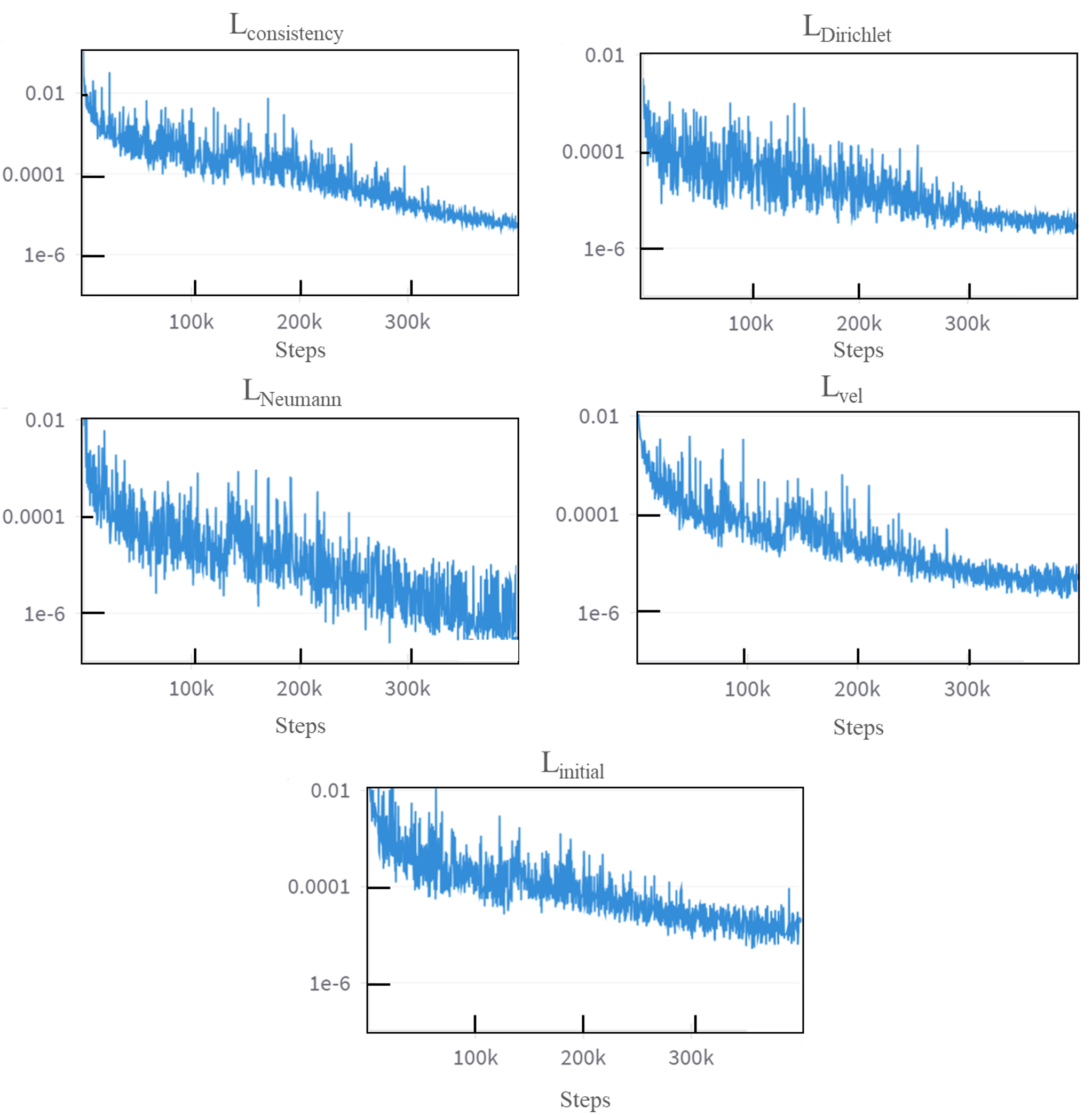}
    \caption{ {The figure shows the value of the loss terms in eq. \eqref{stressLoss} as the network parameters are optimized. We observe that the optimizer uniformly reduces all the loss terms.} }
    \label{fig:loss_values}
\end{figure}

To enforce the Neumann boundary conditions, we use the normal vectors for the wall ($\boldsymbol{n^w} = (l^w, m^w)$) and the outlet ($\boldsymbol{n^o} =  (l^o, m^o)$). We enforce the boundary conditions at the wall using 
\begin{equation}
    L_{wall}(\theta) = \mathbb{E}_{(t^{w},\boldsymbol{x}^{w},\boldsymbol{n}^{w})} \left[|\boldsymbol{\tau}_x(t^{w},\boldsymbol{x}^{w};\theta) {l}^{w} + \boldsymbol{\tau}_y(t^{w},\boldsymbol{x}^{w};\theta) {m}^{w}|^2 \right] ,
\end{equation}
where ({$t^{w},\boldsymbol{x}^{w}$}) is the spatio-temporal point cloud on the walls of the domain. We enforce the boundary condition at the outlet using
\begin{equation}
    L_{outlet}(\theta) =  \mathbb{E}_{(t^{o},\boldsymbol{x}^{o},\boldsymbol{n}^{o})} \left[|\boldsymbol{\tau}_x(t^{o},\boldsymbol{x}^{o};\theta) {l}^{o} + \boldsymbol{\tau}_y(t^{o},\boldsymbol{x}^{o};\theta) {m}^{o}|^2 \right] ,
\end{equation}
where ($t^{o},\boldsymbol{x}^{o}$) is the spatio-temporal point cloud on the outlet of the domain. For the Dirichlet boundary condition at the inlet, we have
\begin{equation}
    L_{inlet}(\theta) =  \mathbb{E}_{(t^{i},\boldsymbol{x}^{i})} \left[\frac{|{\boldsymbol{\tau}}(t^{i},\boldsymbol{x}^{i};\theta)-{\boldsymbol{\tau}}^{i}|^2}{{\sigma_{\boldsymbol{\tau}}}^2} \right] ,
\end{equation}
where ($t^{i},,\boldsymbol{x}^{i}$) are the spatiotemporal point cloud at the inlet of the domain, and $\boldsymbol{\tau}^{i}$ is the stress field at the inlet. We then use eq. \eqref{stressLoss} to define our loss function as 
\begin{equation}\label{stressLoss_stenosis} 
\begin{split} 
L_{stress}(\theta,\phi) = & L_{vel}(\phi) + L_{initial}(\theta) + L_{wall}(\theta) +\\ & L_{outlet}(\theta) + L_{inlet}(\theta) + L_{consistency}(\theta, \phi). 
\end{split} 
\end{equation}
We enforce the Neumann boundary conditions on the wall for the pressure as 
\begin{equation}
    K_{wall}(\kappa) =  \mathbb{E}_{(t^{w},\boldsymbol{x}^{w},\boldsymbol{n}^{w})} \left[|p_x(t^{w},\boldsymbol{x}^{w};\kappa) {l}^{w} + p_y(t^{w},\boldsymbol{x}^{w};\kappa) {m}^{w}|^2 \right],
\end{equation}
and to enforce the Neumann boundary conditions, we use the normal vectors for the inlet ($\boldsymbol{n^i} = (l^i, m^i)$). For the inlet, we enforce 
\begin{equation}
    K_{inlet}(\kappa) =  \mathbb{E}_{(t^{i},\boldsymbol{x}^{i},\boldsymbol{n}^{i})} \left[|p_x(t^{i},\boldsymbol{x}^{i};\kappa) {l}^{i} + p_y(t^{i},\boldsymbol{x}^{i};\kappa) {m}^{i}|^2 \right].
\end{equation}
For the Dirichlet boundary condition at the outlet for the pressure, we have
\begin{equation}
    K_{outlet}(\kappa) =  \mathbb{E}_{(t^{o},\boldsymbol{x}^{o},p^{o})} \left[|{p(t^{o},\boldsymbol{x}^{o};\kappa)-p^{o}}|^2 \right],
\end{equation}
where ${p}^{o}$ is the pressure field at the outlet. We do not divide by the standard deviation of the pressure as the pressure is zero at the outlet in our case. We optimize the parameters $\phi$ and $\kappa$ using eq. \eqref{pLoss} to define the following combined loss
\begin{equation}\label{pLoss_stenosis}
    K_{pressure} (\phi,\kappa) = L_{vel}(\phi) + K_{mom}(\phi,\kappa) + K_{wall}(\kappa) + K_{inlet}(\kappa) + K_{outlet}(\kappa).
\end{equation}
We chose the mini-batch size to be 256 for the spatio-temporal point cloud inside the domain and 64 for all the points on the boundary. Every ten iterations of the Adam optimizer took around 0.45 seconds on a NVIDIA Quadro RTX 8000 GPU.a. We use the following parameters for the Adam optimizer TensorFlow provided: $\beta_1=0.9$, $\beta_2=0.999$, and $\epsilon_1 =1e-07$. Here $\beta_1$ and $\beta_2$ are the exponential decay rates for the first and second momentum estimates, respectively, and $\epsilon_1$ is a small constant for numerical stability. The total running time for the inverse problem is around 25 hours; this includes training all the neural networks and learning the parameters. We use the default parameters for the Adam optimizer provided by tensorflow. All the networks use weight normalization but do not use batch normalization or dropout. {In fig. \ref{fig:loss_values}, we illustrate the reduction of the loss terms in eq. \eqref{stressLoss} throughout the optimization process. It is evident that the optimizer consistently reduces all the loss terms uniformly. }

\bibliographystyle{elsarticle-num}
\bibliography{references}
\end{document}